\pdfoutput=1
\documentclass[a4paper,aip,american,citeautoscript,floatfix,jcp,pdftex,superscriptaddress,%
longbibliography,%
reprint,twocolumn]{revtex4-1}
\usepackage{amsmath,amssymb}
\usepackage{graphicx}
\usepackage[T1]{fontenc}
\usepackage[utf8]{inputenc}
\usepackage{inputenc}
\usepackage{microtype}
\usepackage{textcomp}
\usepackage[textsize=footnotesize]{todonotes}
\usepackage{xspace}
\usepackage{hyperref,hypernat}
\usepackage{units}

\newlength{\figwidth}
\newcommand{\Nuptot}{\ensuremath{\text{N}_{\text{up}}/\text{N}_{\text{tot}}}\xspace}
\newcommand{\Estat}{\ensuremath{\textup{E}_\textup{stat}}\xspace}
\newcommand{\degree}[1]{\ensuremath{#1\,^\circ}\xspace}%
\newcommand{\eg}{e.\,g.}%
\newcommand{\ie}{i.\,e.}%
\newcommand{\cf}{c.\,f.\xspace}%
\newcommand{\kVpcm}{\ensuremath{\text{kV}/\text{cm}}\xspace}
\newcommand{\Vpcm}{\ensuremath{\text{V}/\text{cm}}\xspace}
\newcommand{\Wpcmcm}{\ensuremath{\text{W}/\text{cm}^2}\xspace}
\newcommand{\Icontrol}{\ensuremath{\textup{I}_\textup{control}}\xspace}
\newcommand{\wcmcm}{\Wpcmcm}
\newcommand{\kvcm}{\kVpcm}
\newcommand{\vcm}{\Vpcm}

\newcommand{\cfeldesy}{\affiliation{Center for Free-Electron Laser Science, Deutsches
      Elektronen-Synchrotron DESY, Notkestrasse 85, 22607 Hamburg, Germany}}%
\newcommand{\uhhcui}{\affiliation{The Hamburg Center for Ultrafast Imaging, Universität Hamburg,
      Luruper Chaussee 149, 22761 Hamburg, Germany}}%
\newcommand{\uhhphys}{\affiliation{Department of Physics, Universität Hamburg, Luruper Chaussee 149,
      22761 Hamburg, Germany}}%
\newcommand{\granada}{\affiliation{Instituto Carlos I de F\'{\i}sica Te\'orica y Computacional and
      Departamento de F\'{\i}sica At\'omica, Molecular y Nuclear, Universidad de Granada, 18071
      Granada, Spain}}%

\begin{document}
\title{Time-dependent analysis of the mixed-field orientation of molecules without rotational
   symmetry}%
\author{Linda V. Thesing}\cfeldesy%
\author{Jochen Küpper}%
\email{jochen.kuepper@cfel.de}%
\homepage[\newline\hspace*{0.8em}]{https://www.controlled-molecule-imaging.org}%
\cfeldesy\uhhcui\uhhphys%
\author{Rosario Gonz{\'a}lez-F{\'e}rez}%
\email{rogonzal@ugr.es}%
\granada%
\date{\today}%
\begin{abstract}\noindent%
   We present a theoretical study of the mixed-field orientation of molecules without rotational
   symmetry. The time-dependent one-dimensional and three-dimensional orientation of a thermal
   ensemble of 6-chloropyridazine-3-carbonitrile molecules in combined linearly or elliptically
   polarized laser fields and tilted dc electric fields is computed. The results are in good
   agreement with recent experimental results of one-dimensional orientation for weak dc electric
   fields [\emph{J.\ Chem.\ Phys.}~\textbf{139}, 234313 (2013)]. Moreover, they predict that using
   elliptically polarized laser fields or strong dc fields three-dimensional orientation is obtained.
   The field-dressed dynamics of excited
   rotational states is characterized by highly non-adiabatic effects. We analyze the sources of
   these non-adiabatic effects and investigate their impact on the mixed-field orientation for
   different field configurations in mixed-field-orientation experiments.
\end{abstract}
\maketitle%
\noindent%

\section{Introduction}%
Directional control over complex molecules, \ie, their alignment and orientation in laboratory
space, is strongly sought after for many applications, not the least in the quest for the recording
of so-called molecular movies of (bio)chemical dynamics~\cite{Spence:PRL92:198102, Barty:ARPC64:415,
   Miller:ARPC65:583}. A detailed understanding of the underlying control mechanisms and the related
rotational dynamics is necessary to guide experiments as well as to support the analysis of the
actual imaging experiments, especially when no good experimental observables of the degree of
alignment and orientation are available anymore.

Even for present day experiments with moderately small molecules the angular control strongly
improves or simply enables the observation of steric effects in chemical
reactions~\cite{Brooks:Science193:11, Rakitzis:Science303:1852} or in imaging experiments utilizing
photoelectron angular distributions~\cite{Meckel:Science320:1478, Bisgaard:Science323:1464,
   Holmegaard:NatPhys6:428}, high-order-harmonic-generation
spectroscopy~\cite{Itatani:Nature432:867, Vozzi:NatPhys7:822, Kraus:PRL113:023001}, and X-ray and
electron diffractive imaging~\cite{Hensley:PRL109:133202, Kuepper:PRL112:083002,
   Yang:NatComm7:11232}. While these experiments typically have been performed for simpler molecules
so far, molecules of the complexity of 6-chloropyridazine-3-carbonitrile (CPC) are within reach.

The generation as well as the basic concepts of alignment and orientation have been described
before\cite{Loesch:JCP93:4779, Friedrich:JCP111:6157, Stapelfeldt:RMP75:543}. Here, we focus on
mixed-field orientation, which utilizes the combined action of a strong laser field with a dc
electric field~\cite{Friedrich:JCP111:6157, Holmegaard:PRL102:023001, Ghafur:NatPhys5:289,
   Nielsen:PRL108:193001, Trippel:PRL114:103003}. This approach was experimentally shown to allow
for three-dimensional (3D) alignment and one-dimensional (1D) orientation of the prototypical
complex molecule CPC~\cite{Hansen:JCP139:234313}, but the employed adiabatic analysis of the quantum
dynamics could not reproduce the experimental findings~\cite{Omiste:PCCP13:18815,
   Hansen:JCP139:234313}.

Here, we set out to accurately theoretically describe the rotational dynamics of a thermal ensemble
of CPC under the combined action of laser pulses and dc electric field. We investigate the influence
of the dc field strength as well as the angle between the fields as the ac field strength changes
during the turn-on of the linearly and elliptically polarized laser fields. Our findings demonstrate
that the rotational dynamics is very complex, and can be quantitatively described as a solution of
the time-dependent Schrödinger equation.

\section{Theoretical Description}
We consider a planar, polar, asymmetric top molecule with a polarizability tensor that is diagonal
in the principle-axes-of-inertia system and an electric dipole moment (EDM) that is not parallel to
any principle axis of inertia. We investigate the rotational dynamics of such a molecule in combined
non-resonant laser and dc electric fields where the laser field is either linearly or elliptically
polarized. The (major) polarization axis of the laser defines the $Z$-axis of the laboratory-fixed
frame (LFF) ($X,Y,Z$) and for an elliptically polarized laser, the LFF $Y$-axis is defined by the
minor polarization axis. The static field lies in the $YZ$-plane forming an angle $\beta$ with the
$Z$-axis. The molecule-fixed frame (MFF) ($x,y,z$) is defined by the principle axes of inertia in
such a way that the rotational constants satisfy $B_z>B_y>B_x$. The MFF is related to the LFF by the
Euler angles ($\phi,\theta,\chi$).

Within the rigid rotor approximation, the Hamiltonian of the system is
\begin{equation}
   H(t) = J^2_x B_x+ J^2_y B_y +J^2_z B_z + H_\text{stat} + H_\text{laser}(t)
   \label{eq:hamiltonian}
\end{equation}
where $J_i$ are the components of the angular momentum with respect to the $i$-axis,
$i\in\lbrace{}x,y,z\rbrace$. The interaction of the molecule with the static electric field reads
\begin{equation}
   H_\text{stat}= - {\textbf{E}_\textup{stat}}\xspace \cdot \boldsymbol{\mu} = - \Estat \mu_z \cos \theta_{sz} - \Estat \mu_y \cos \theta_{sy}
   \label{eq:interaction_static}
\end{equation}
with the dc field strength \Estat, the components $\mu_y$ and $\mu_z$ of the EDM $\boldsymbol{\mu}$
and the angles $\theta_{sz}$ and $\theta_{sy}$ between the static electric field and the MFF $z$- and
$y$-axes, respectively (see ref.~\onlinecite{Hansen:JCP139:234313} for their relations to the Euler angles).
The interaction of the molecule with the non-resonant laser field reads
\begin{align}
  H_\text{laser}(t) =
  &-\frac{I_{ZZ}(t)}{2c\varepsilon_0}\left(\alpha^{zx}\cos^2\theta_{Zz}+\alpha^{yx}\cos^2\theta_{Zy}\right)\label{eq:interaction_laser}\\
  &-\frac{I_{YY}(t)}{2c\varepsilon_0}\left(\alpha^{zx}\cos^2\theta_{Yz}+\alpha^{yx}\cos^2\theta_{Yy}\right)\nonumber
\end{align}
where $I_{ZZ}(t)$ and $I_{YY}(t)$ are the time-dependent intensities along the major and minor
laser-field polarization axes, respectively. For linear polarization, $I_{YY}(t)=0$ and the total
intensity is given by $I(t)=I_{ZZ}(t)$; for elliptical polarization $I(t)=I_{ZZ}(t)+I_{YY}(t)$ and
we assume $I_{ZZ}(t)=3I_{YY}(t)$. In~\eqref{eq:interaction_laser},
$\alpha^{ij}=\alpha_{ii}-\alpha_{jj}$, where $\alpha_{ii}$ are the diagonal elements of the
polarizability tensor with $i,j\in\lbrace{x,y,z}\rbrace$. $\theta_{Pq}$ is the angle between the LFF
$P$-axis and the MFF $q$-axis~\cite{Hansen:JCP139:234313}.

To investigate the rotational dynamics of the CPC molecule in combined ac and dc fields, we solve
the time-dependent Schrödinger equation (TDSE) associated with the
Hamiltonian~\eqref{eq:hamiltonian}. For the angular coordinates, we employ a basis set expansion of
the wave function using linear combinations of field-free symmetric rotor states, \ie, Wigner
functions~\cite{Zare:AngularMomentum}, which respect the symmetries of the Hamiltonian. In the case
of tilted fields with $\beta \neq \degree{90}$, the remaining symmetry operations of the
Hamiltonian~\eqref{eq:hamiltonian} are the identity $E$ and the reflection $\sigma_{YZ}$ on the
$YZ$-plane, which contains the dc and ac electric fields, implying two irreducible representations.
The basis used for each irreducible representation has been described
elsewhere~\cite{Omiste:JCP135:064310}. The time propagation is carried out using the short iterative
Lanczos method~\cite{Leforestier:JCOP94:59, Beck:PhysRep324:1}. In our calculations, the dynamics
during the turn-on of the static electric field is assumed to be adiabatic and the dc field-dressed
states are taken as the initial states of the time-propagation. For the adiabatic labeling of the
time-dependent states, the static electric field is first turned on parallel to the LFF $Z$-axis
and, thereafter, rotated by an angle $\beta$.

\begin{figure}
   \centering%
   \includegraphics[width=0.66\linewidth]{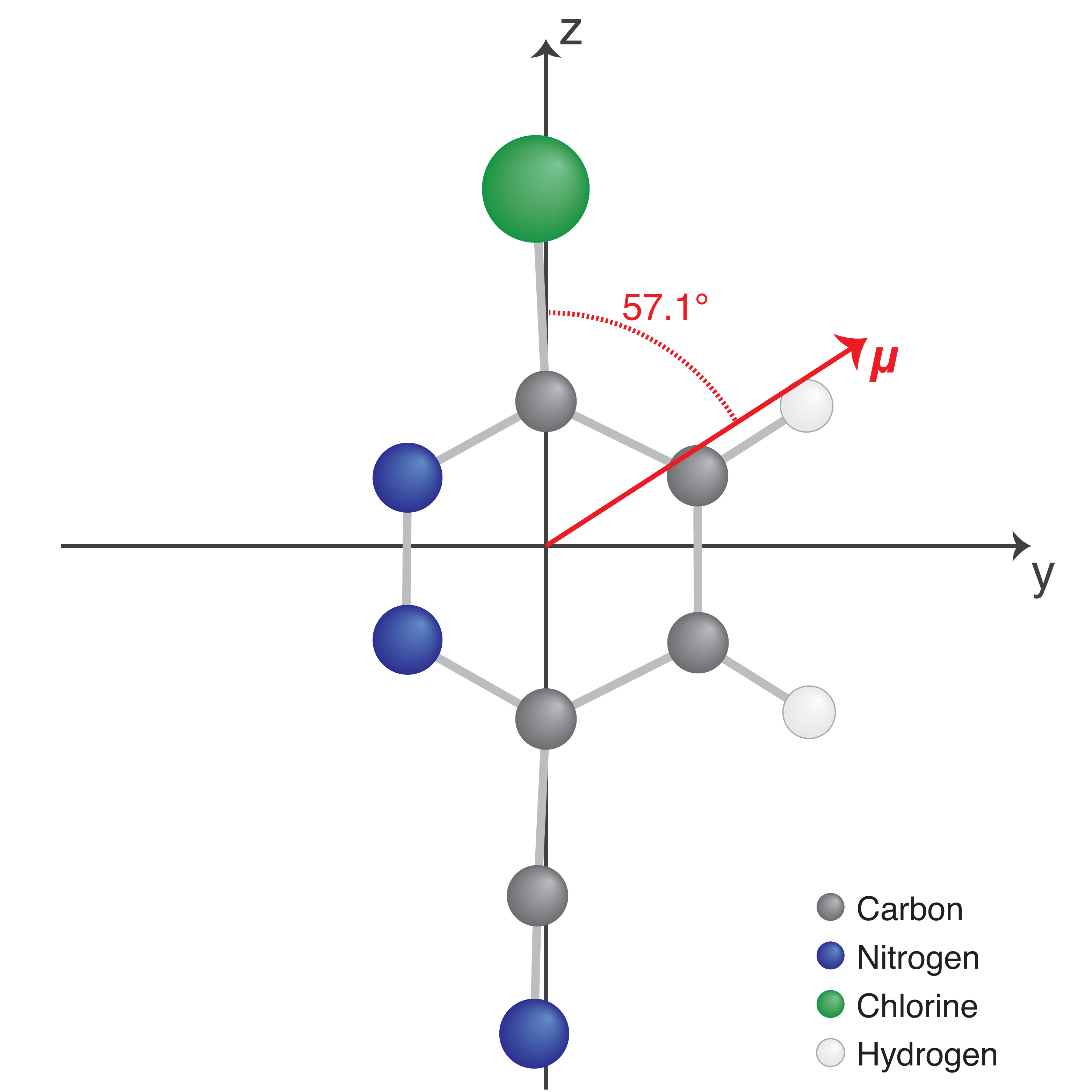}
   \caption{Sketch of the structure of the CPC molecule. The $x$-axis is perpendicular to the figure
      plane.}
   \label{fig:CPC_structure}
\end{figure}

This work focuses on the CPC molecule~\cite{Hansen:JCP139:234313}, which has an EDM of $\mu=5.21$~D
that forms an angle of \degree{57.1} with the most polarizable axis (MPA) of the molecule,
see~\autoref{fig:CPC_structure}. The components of the EDM as well as the rotational constants and
polarizability components of CPC are listed in~\autoref{table:mol_parameters}.
\begin{table}
   \begin{tabular*}{\linewidth}{@{\extracolsep{\fill}}l c c c}
     \hline \hline
     $i$ & $B_i$ (MHz)& $\alpha_{ii}$ ($10^{-3}~\text{nm}^3$) & $\mu_i$ (D) \\
     \hline
     $x$ & 639.708 & 7.88   & 0     \\
     $y$ & 717.422 & 11.98 & 4.37 \\
     $z$ & 5905.0   & 22.24 & 2.83 \\
     \hline \hline
   \end{tabular*}
   \caption{Rotational constants, polarizability components and components of the EDM
      of the CPC molecule~\cite{Hansen:JCP139:234313}. In the calculation of the rotational
      constants, the $^{35}$Cl isotope was considered.
   }
\label{table:mol_parameters}
\end{table}
We use a field configuration equivalent to the experimental one~\cite{Hansen:JCP139:234313}, \ie, a
Gaussian laser pulse with a peak intensity of $\Icontrol=8.0\times10^{11}~\wcmcm$ and a full-width
at half maximum (FWHM) of 10~ns, and a tilted dc electric field with a field strength of 571~\vcm.

In the experiment~\cite{Hansen:JCP139:234313}, an inhomogeneous electric field was used to deflect
the molecules according to the effective dipole moments of their rotational
states~\cite{Filsinger:JCP131:064309, Chang:IRPC34:557}. The dispersed molecular beam then entered a
velocity map imaging (VMI) spectrometer where it was crossed by the alignment and probe laser
pulses. A state-selection of the sample was carried out by focusing the laser beams on the most
deflected molecules, which are the ones in the lowest-energy rotational
states~\cite{Chang:IRPC34:557}. The alignment and orientation of the CPC molecules was measured by
Coulomb explosion imaging of ionic fragments onto a screen perpendicular to the dc electric field.
The measured orientation corresponds to the average orientation of the molecules occupying different
rotational states in the state-selected molecular ensemble.

The degree of orientation is experimentally quantified by the ratio \Nuptot of Cl$^+$ ions that are
detected in the upper half of the screen to the total number of detected ions. In our calculations,
we determine this orientation ratio by projecting the 3D probability density onto a 2D screen
perpendicular to the dc electric field~\cite{Omiste:PCCP13:18815}. Here, we take into account the
probe selectivity for a probe pulse that is linearly polarized parallel to the screen. The volume
effect due to the spatial intensity profiles of the alignment and probe lasers is not included in
this work. We mimic the state-selection, which results in a non-thermal population distribution,
by using a thermal ensemble at a temperature of 200~mK,
which is lower than the (1~K) rotational temperature of the non-deflected molecular sample and was
found to describe the non-thermal experimental ensemble well. We solve the TDSE for each state
$\left|\psi_\gamma\right>$ in the thermal ensemble and determine the thermal orientation ratio as
\begin{equation}
   \frac{\text{N}_\text{up}}{\text{N}_\text{tot}}
   = \sum_\gamma w^T_{\gamma} \frac{\text{N}_\text{up}^{\gamma}}{\text{N}_\text{tot}^{\gamma}}
   \label{thermal_average}
\end{equation}
with the thermal weights according to a Boltzmann distribution
\begin{equation}
   w^T_{\gamma} = \frac{e^{-E_{\gamma}/k_B T}}{\sum_\gamma e^{-E_{\gamma}/k_B T}}
   \label{thermal_weights}
\end{equation}
with the field-free energy $E_{\gamma}$ of the state $\left|\psi_\gamma\right>$. For the thermal
ensemble at 200~mK, we take into account the 50 lowest lying rotational states of each irreducible
representation, \ie, 100 rotational states are included in the sums \eqref{thermal_average}
and~\eqref{thermal_weights}.

\section{Orientation of a thermal ensemble and comparison to experimental results}
\begin{figure}
   \includegraphics[width=\linewidth]{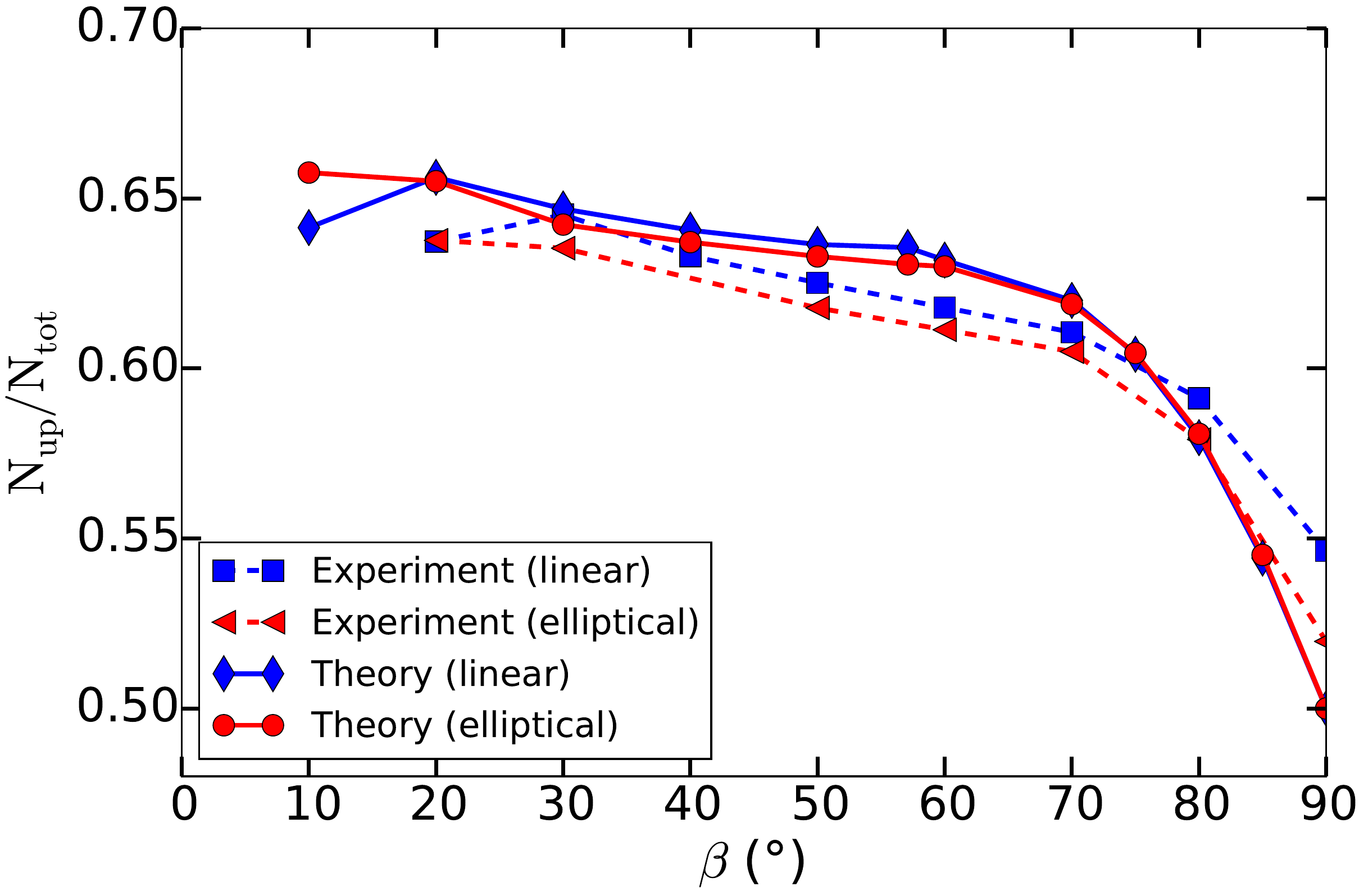}
   \caption{The orientation ratio \Nuptot at the peak of the laser pulse,
      $\Icontrol=8.0\times10^{11}~\wcmcm$, as a function of $\beta$ for linearly and elliptically
      polarized pulses, computed for a thermal ensemble at $T=200$~mK (red diamonds and blue
      circles, respectively) and experimental results for a state-selected sample (red squares and
      blue triangles), see Fig.~5 from ref.~\onlinecite{Hansen:JCP139:234313}. The dc field strength
      is 571~\vcm.}
   \label{fig:comparison_experiment}
\end{figure}
\autoref{fig:comparison_experiment} shows the theoretical orientation for a thermal ensemble at
$T=200$~mK at the control laser peak as a function of the angle $\beta$ between the dc field and the
LFF $Z$-axis for linearly and elliptically polarized laser fields. These results are in very good
agreement with the experimental orientation ratios for the state-selected molecular sample, which
are reproduced from ref.~\onlinecite{Hansen:JCP139:234313}. We observe similar results for the
linearly and elliptically polarized laser fields: The orientation of the thermal ensemble increases
smoothly as the static field is rotated towards the (major) polarization axis of the laser, and
reaches values of $\Nuptot\sim0.65$ for small $\beta$. For linear polarization, \Nuptot decreases
for $\beta=\degree{10}$ due to the employed geometry of the experimental setup, which cannot
correctly measure the orientation for parallel fields~\cite{Nielsen:PRL108:193001}. Overall, our
time-dependent description of the mixed-field orientation of state-selected CPC reproduces the
features of the experimentally observed behavior well. In particular, we find a good agreement
between the experimental and theoretical orientation ratio for $\beta\lesssim\degree{70}$. For
$\beta=\degree{90}$, the experimental ensemble shows a small orientation which contradicts the
theoretical prediction of no orientation for perpendicular fields. This could be due to experimental
imperfections such as misalignment of the dc field, the polarization axis and detector screen or due
to the influence of the probe laser. We point out that the smooth behavior of \Nuptot is a
consequence of the ensemble average. The orientation of an individual state might strongly depend on
$\beta$ due to highly non-adiabatic effects appearing in the rotational dynamics,
see~\autoref{subsection_dynamics} below. These non-adiabatic effects are masked by the ensemble
average.

\begin{figure}
   \includegraphics[width=\linewidth]{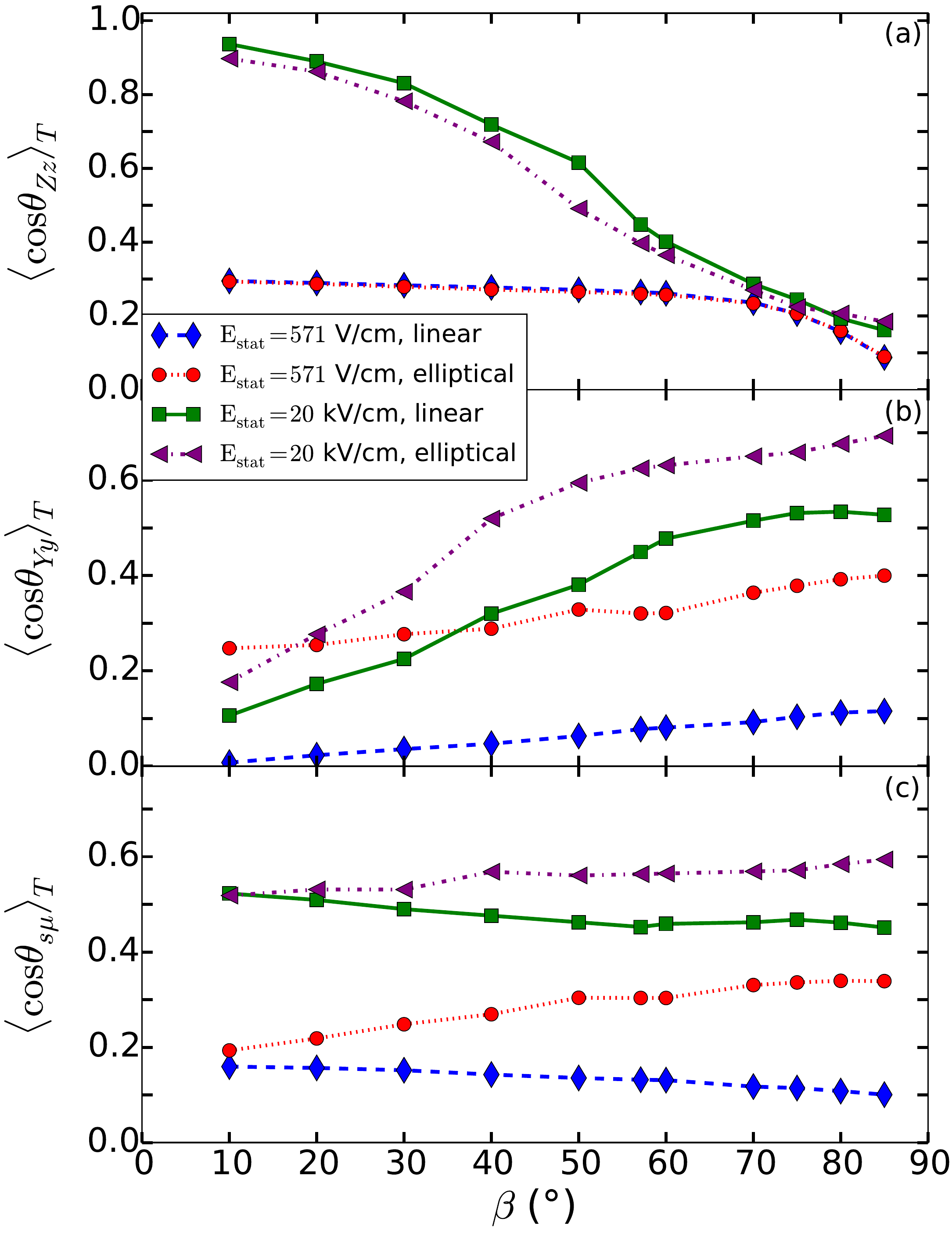}
   \caption{For a thermal ensemble at $T=200$~mK, the expectation values (a)
      $\langle\cos\theta_{Zz}\rangle_T$, (b) $\langle\cos\theta_{Yy}\rangle_T$ and (c)
      $\langle\cos\theta_{s\mu}\rangle_T$ at the laser peak intensity as a function of $\beta$ for a
      linearly polarized ac field and with $\Estat=571$~\vcm (blue diamonds) and $\Estat=20$~\kvcm
      (green squares) as well as for an elliptically polarized ac field with $\Estat=571$~\vcm (red
      circles) and $\Estat=20$~\kvcm (purple triangles).}
   \label{fig:orientation_thermal}
\end{figure}
\autoref{fig:orientation_thermal} shows the calculated expectation values of the orientation cosines
along the LFF $Z$ and $Y$-axes as well as the dc field direction for the same thermal sample and
field configurations. The ratio \Nuptot plotted in~\autoref{fig:comparison_experiment} quantifies
the orientation of the MFF $z$-axis along the major polarization axis of the laser, providing
information about 1D orientation, which is also characterized by $\langle\cos\theta_{Zz}\rangle_T$
shown in~\autoref{fig:orientation_thermal}~(a). To investigate the 3D orientation, we additionally
consider the expectation value $\langle\cos\theta_{Yy}\rangle_T$, shown
in~\autoref{fig:orientation_thermal}~(b), which measures the orientation of the MFF $y$-axis along
the LFF $Y$-axis. The orientation of the EDM along the dc field direction is quantified by
$\langle\cos\theta_{s\mu}\rangle_T$, presented in~\autoref{fig:orientation_thermal}~(c).

\begin{figure}
   \includegraphics[width=\linewidth]{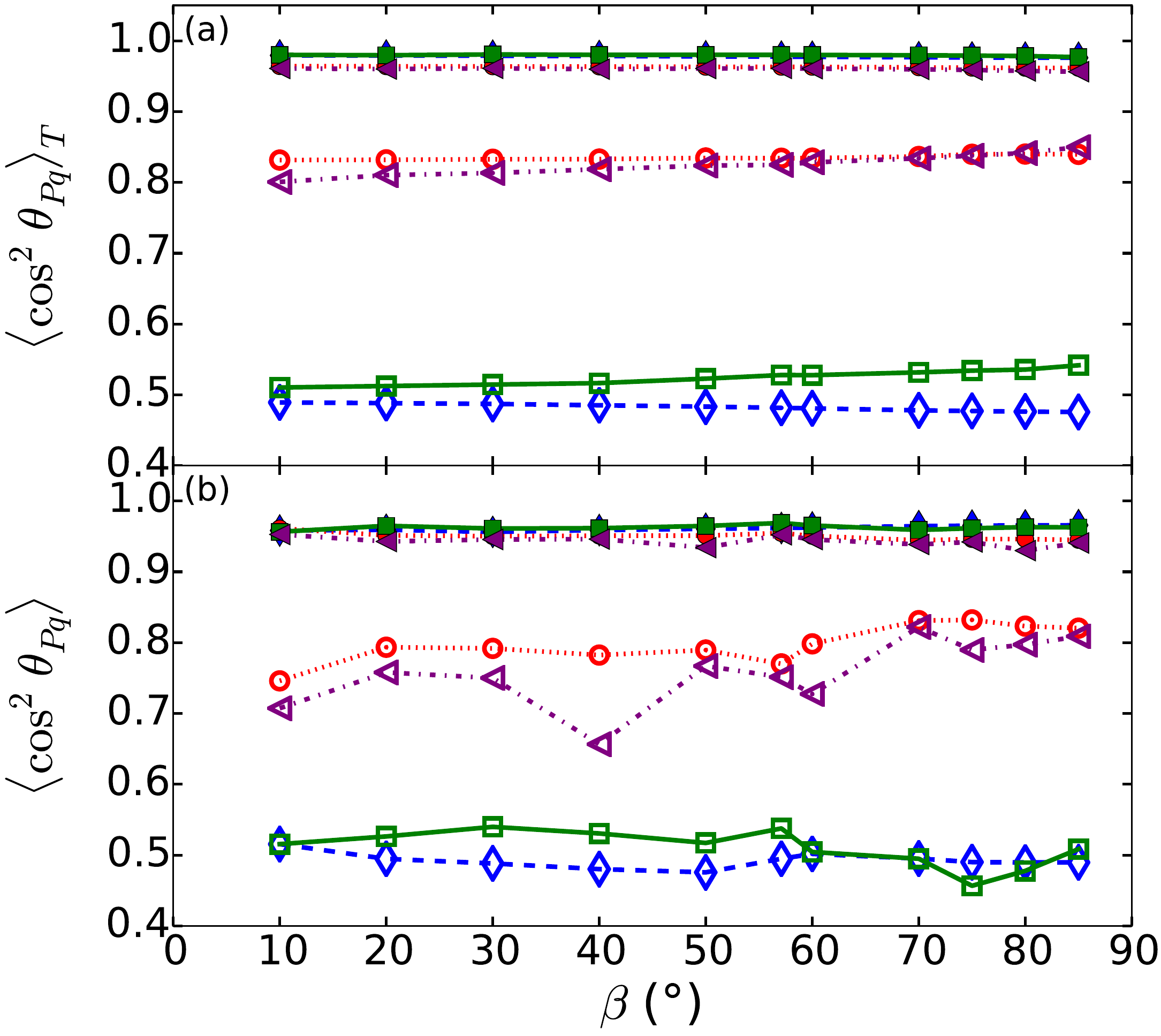}
   \caption{For (a) a thermal ensemble at $T=200$~mK, and (b) the state $|2_{12}1\rangle_\text{t}$,
      the expectation values $\langle\cos^2\theta_{Zz}\rangle$ (filled symbols) and
      $\langle\cos^2\theta_{Yy}\rangle$ (empty symbols) at the laser peak intensity as a function of
      $\beta$ for a linearly polarized ac field and with $\Estat=571$~\vcm (blue diamonds) and
      $\Estat=20$~\kvcm (green squares) as well as for an elliptically polarized ac field with
      $\Estat=571$~\vcm (red circles) and $\Estat=20$~\kvcm (purple triangles). The alignment along
      the LFF $Z$-axis is practically independent of all parameters with values of
      $\langle\cos^2\theta_{Zz}\rangle_T\approx0.97$ and
      $\langle\cos^2\theta_{Zz}\rangle\approx0.95$ for the thermal ensemble and the
      $|2_{12}1\rangle_\text{t}$ state, respectively.}
   \label{fig:alignment}
\end{figure}
For the linearly polarized laser field, the thermal sample is strongly aligned along the $Z$-axis, but
not aligned along the $Y$-axis, as presented in~\autoref{fig:alignment}. Strong 3D alignment
is obtained with the elliptically polarized laser field, yielding
$\langle\cos^2\theta_{Zz}\rangle_T\approx0.95$ and $\langle\cos^2\theta_{Yy}\rangle_T\approx0.83$ for
all $\beta$. The alignment cosine $\langle\cos^2\theta_{Yy}\rangle_T$ shows a weak dependence on dc
field and on the angle $\beta$.

The orientation of the thermal ensemble, \autoref{fig:orientation_thermal}~(a), shows a smooth
behavior, with $\langle\cos\theta_{Zz}\rangle_T$ increasing from 0.09 to 0.29 for
$\beta=\degree{85}$ and $\beta=\degree{10}$, respectively. This increase is not reproduced within
the adiabatic picture, which predicts a $\beta$-independent orientation for a weak static electric
field~\cite{Hansen:JCP139:234313}. The small orientation for $\beta$ close to \degree{90}
in~\autoref{fig:orientation_thermal}, obtained from the time-dependent results, can be explained by
population transfer within pendular doublets formed as the laser intensity
rises~\cite{Nielsen:PRL108:193001, Omiste:PRA86:043437, Omiste:PRA88:033416, Trippel:PRL114:103003}.
At this field configuration, the oriented and antioriented adiabatic states within these
near-degenerate pairs have similar contributions to the field-dressed dynamics reducing the
orientation of the ensemble.

We now turn to the influence of a strong dc electric field on the orientation. For a dc field of
20~\kvcm, the thermal orientation $\langle\cos\theta_{Zz}\rangle_T$ rapidly increases as the angle
between the fields decreases. For $\beta=\degree{10}$, we find a strong orientation with
$\langle\cos\theta_{Zz}\rangle_T=0.94$ and $\langle\cos\theta_{Zz}\rangle_T=0.90$ for linearly and
elliptically polarized pulses, respectively. The enhancement of the orientation can be rationalized
in terms of the rotational dynamics of each state in the thermal ensemble, \ie, the contributions of
different instantaneous eigenstates of the Hamiltoninan~\eqref{eq:hamiltonian} to the time-dependent
wave function, see~\autoref{subsection_dynamics}. For low laser intensities, the interaction with
the dc field is dominant, inducing a brute-force orientation of the EDM along the dc field
direction. If $\beta$ is small, this brute-force orientation implies a moderate orientation of the
MFF $z$-axis along the LFF $Z$-axis, due to the angle between the EDM and the MFF $z$-axis. As the
laser intensity increases and becomes the dominant interaction, the strong dc field leads to
preferred contributions of oriented pendular states to the time-dependent wave functions, which
results in an enhancement of the orientation of the thermal ensemble.

Considering the orientation of the MFF $y$-axis along the LFF $Y$-axis, shown in
\autoref{fig:orientation_thermal}~(b), the thermal ensemble is very weakly oriented for a linearly
polarized pulse and $\Estat=571$~\vcm, reaching a maximum value of
$\langle\cos\theta_{Yy}\rangle_T=0.12$ at $\beta=\degree{85}$. However, a strong dc electric field
induces a brute-force orientation of the EDM, and thus, a moderate orientation of the molecular
$y$-axis, along the dc field direction before the laser pulse is turned on. This orientation of the
MFF $y$-axis is maintained to some extend even in the presence of the laser field with
$\langle\cos\theta_{Yy}\rangle_T$ increasing from 0.11 to 0.53 for $\beta=\degree{10}$ and
$\beta=\degree{85}$, respectively, at the peak of the linearly-polarized laser pulse. Thus, a strong
dc field combined with a linearly polarized ac field induces a significant 3D orientation for
intermediate $\beta$. In contrast, even a weak dc field induces a moderate 3D orientation when
combined with an elliptically polarized laser field. Here, the orientation along the $Y$-axis
monotonically increases from $\langle\cos\theta_{Yy}\rangle_T=0.25$ to
$\langle\cos\theta_{Yy}\rangle_T=0.40$ for $\beta=\degree{10}$ to $\beta=\degree{85}$, respectively.
For $\Estat=20$~\kvcm, we find an enhancement of $\langle\cos\theta_{Yy}\rangle_T$, similar to the
behavior of $\langle\cos\theta_{Zz}\rangle_T$ for strong dc fields, but for increasing $\beta$. If
$\beta$ is small, the influence of the static field along the $Y$-axis is not strong enough to, in
general, achieve preferred contributions of oriented pendular states to the rotational dynamics of
excited states in the thermal ensemble. As a result, for $\beta=\degree{10}$ we encounter a few
low-lying excited states in the molecular ensemble that are not oriented or even strongly
antioriented along the $Y$-axis at the peak intensity. Compared to the weak dc field,
$\langle\cos\theta_{Yy}\rangle_T$ is reduced at this angle. This effect disappears if the
temperature is increased and more excited states contribute to the thermal ensemble.

\begin{figure}
   \includegraphics[width=\linewidth]{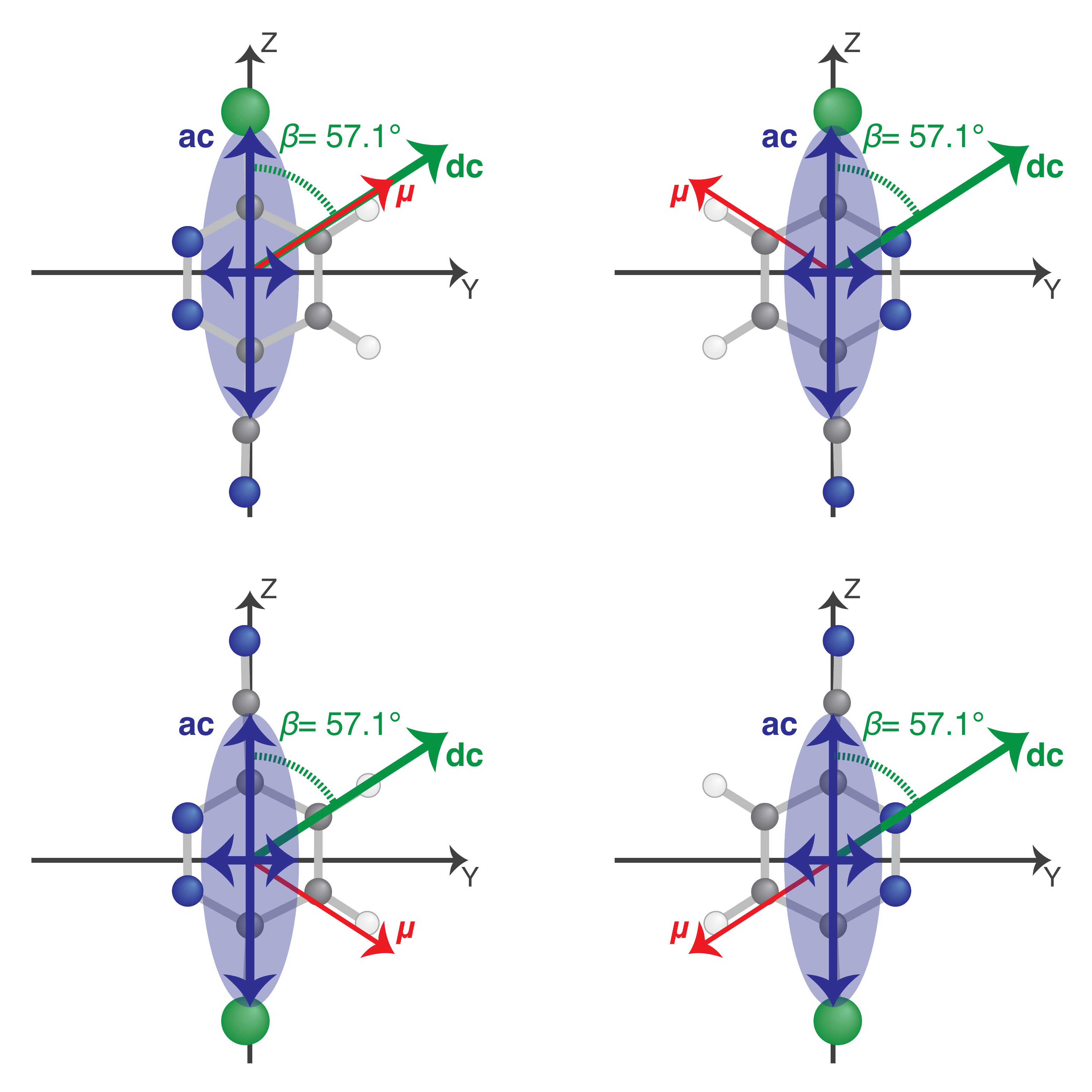}
   \caption{For an elliptically polarized laser field, the four possible orientations of the 3D
      aligned pendular states in the adiabatic picture.}
   \label{fig:orientations_57}
\end{figure}
Regarding the behavior of $\langle\cos\theta_{s\mu}\rangle_T$
in~\autoref{fig:orientation_thermal}~(c),\footnote{This expectation value is given by
   $\protect\langle\cos\theta_{s\mu}\protect\rangle =
   \cos(\degree{57.1})\protect\langle\cos\theta_{sz}\protect\rangle+
   \sin(\degree{57.1})\protect\langle\cos\theta_{sy}\protect\rangle$.} the orientation of the EDM
along the dc field direction shows a weak dependence on $\beta$.
For small angles between the ac and
dc fields, $\langle\cos\theta_{s\mu}\rangle_T$ is approximately given by
$\cos(\degree{57.1}) \langle\cos\theta_{Zz}\rangle_T$, since for these small angles it holds
$\langle\cos\theta_{sz}\rangle_T\approx\langle\cos\theta_{Zz}\rangle_T$. Thus, for both
polarizations we find similar values of $\langle\cos\theta_{s\mu}\rangle_T\approx0.18$ and
\mbox{$\langle\cos\theta_{s\mu}\rangle_T\approx0.5$} for the weak and strong dc fields,
respectively. For $\beta=\degree{85}$, the orientation of the EDM along the dc field direction is
dominated by the behavior of $\langle\cos\theta_{Yy}\rangle_T$, and as a consequence,
$\langle\cos\theta_{s\mu}\rangle_T$ is larger for the elliptically polarized laser field than for
the linearly polarized laser field. The decreasing (increasing) behavior of
$\langle\cos\theta_{s\mu}\rangle_T$ versus $\beta$ for linear (elliptical) polarization is due to
the larger contribution of $\langle\cos\theta_{sy}\rangle_T$, since
$\cos(\degree{57.1})<\sin(\degree{57.1})$. In addition, for the linear polarized case we find that
$\langle\cos\theta_{Yy}\rangle_T$ for $\beta=\degree{85}$ is smaller than
$\langle\cos\theta_{Zz}\rangle_T$ for $\beta=\degree{10}$ for both dc field strengths. For
$\beta=\degree{57.1}$, the angles between the ac and dc fields and between the EDM and the MPA
coincide, and a maximum of $\langle\cos\theta_{s\mu}\rangle_T$ could be expected for the 3D aligned
sample in an elliptically polarized laser field. However, this is not the case since the orientation
of the EDM is obtained from the average over the four possible orientations of the adiabatic pendular
states, shown in~\autoref{fig:orientations_57}, which contribute
to the time-dependent wave function of each state in the thermal ensemble.

To summarize, a significant 3D orientation is obtained for intermediate angles $\beta$, a strong dc
field and both, linear and elliptical, laser polarizations. The degree of orientation of the MFF
$y$-axis along the LFF $Y$-axis is larger for an elliptically polarized pulse than for a linearly
polarized one, whereas the orientation of the MFF $z$-axis along the LFF $Z$-axis is similar in both
cases. This implies that to achieve a large 3D orientation an elliptically polarized laser is
recommended. Nevertheless, we point out that the naive picture of achieving 3D orientation by fixing
one molecular axis, namely the MPA, with a linearly polarized laser field and a second axis, the
EDM, with a dc electric field does (only) work in the limit of strong laser and strong dc electric
fields.

\section{Field-dressed dynamics}
\label{subsection_dynamics}
\begin{figure}
   \includegraphics[width=\linewidth]{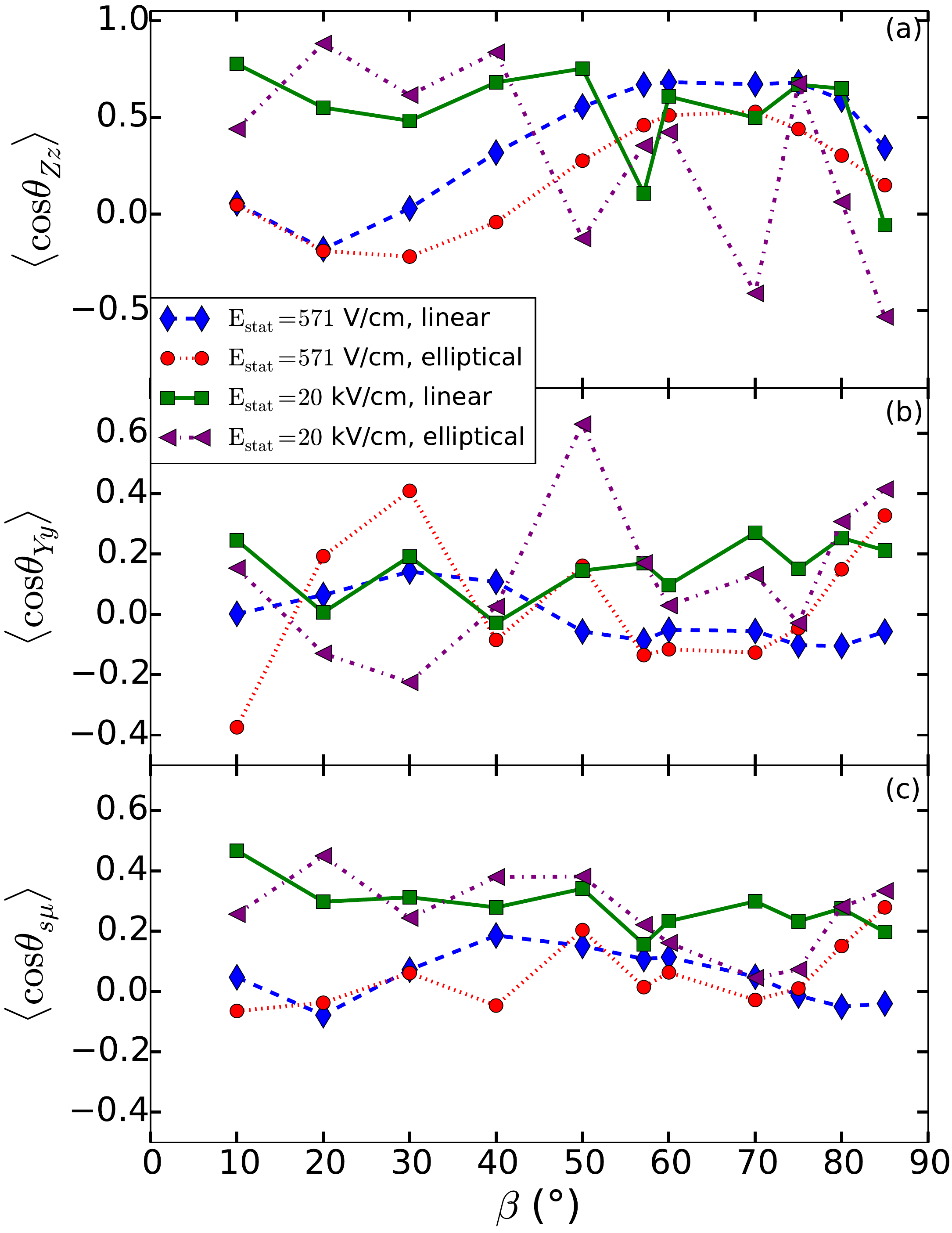}
   \caption{The expectation values (a) $\langle\cos\theta_{Zz}\rangle$, (b)
      $\langle\cos\theta_{Yy}\rangle$ and (c) $\langle\cos\theta_{s\mu}\rangle$ for the state
      $|2_{12}1\rangle_\text{t}$ at the peak intensity as a function of $\beta$ for a linearly
      polarized ac field and dc field strength $\Estat=571$~\vcm (blue diamonds) and
      $\Estat=20$~\kvcm (green squares) as well as for an elliptically polarized ac field and dc
      field strength $\Estat=571$~\vcm (red circles) and $\Estat=20$~\kvcm (purple triangles).}
   \label{fig:orientation_2121}
\end{figure}

To analyze the dynamics of individual excited states\footnote{We use the notations
   $|J_{K_a K_c}M\protect\rangle_\text{t}$ and $|J_{K_a K_c}M\protect\rangle_\text{p}$ for the
   time-dependent wave function and the adiabatic pendular states, respectively. Here, we only
   analyze the dynamics for states with even parity under reflection on the plane containing the ac
   and dc fields.} for the different field configurations in detail, we choose the state
$|2_{12}1\rangle_\text{t}$ as a prototypical example. Energetically, it is the 14th and 19th excited
state of the representation with even parity under the reflection $\sigma_{YZ}$ for
$\Estat=571$~\vcm and $\Estat=20$~\kvcm, respectively. In~\autoref{fig:orientation_2121} the
orientation cosines $\langle\cos\theta_{Zz}\rangle$, $\langle\cos\theta_{Yy}\rangle$ and
$\langle\cos\theta_{s\mu}\rangle$ at the peak intensity are presented as a function of $\beta$.
The smooth $\beta$-dependence found for the orientation of the thermal ensemble is not reflected in
the final orientation of the state $|2_{12}1\rangle_\text{t}$. For instance, depending on the angle
$\beta$, this state can be strongly 1D or 3D oriented, antioriented, or not oriented. However,
certain features of the orientation of the thermal average can also be observed for the state
$|2_{12}1\rangle_\text{t}$ as well as for most other states contributing to the ensemble average at
200~mK. In particular, the significant orientation along the $Z$-axis for a strong dc field and
small $\beta$ and the decreasing orientation along the $Z$-axis for a weak dc field in the region
$\degree{75}\lesssim\beta<\degree{90}$. However, the exact orientation for a specific field
configuration can only be determined from time-dependent calculations. In contrast, the alignment
along the the $Z$-axis of the state $|2_{12}1\rangle_\text{t}$, shown
in~\autoref{fig:alignment}~(b), does not depend on $\beta$ and resembles the alignment of the
thermal sample. The alignment cosine $\langle\cos^2\theta_{Yy}\rangle$ shows a weak dependence on
the angle between the ac and dc fields due to contributions of highly excited pendular states that
are weakly aligned along the $Y$-axis.

The rotational dynamics shows highly non-adiabatic effects due to $M$-manifold splitting, avoided
crossings and the formation of pendular doublets and quadruplets~\cite{Omiste:PRA88:033416,
   Omiste:PRA94:063408}. The importance of each of these effects depends strongly on the field
configuration, \ie, the ac and dc field strengths, the angle $\beta$ between the fields, and the
polarization and temporal profile of the laser pulse. In this section, we investigate the influence
of these non-adiabatic phenomena on the mixed-field orientation. For illustration, the time
evolution of the rotational dynamics of the $|2_{12}1\rangle_\text{t}$ state and the orientation
cosines of the adiabatic pendular states are shown in movies provided in the supplementary
material (see~\autoref{subsection_SI}).

\subsection{Linearly polarized laser}
\label{subsection_dynamics_lin}
\begin{figure}
   \includegraphics[width=\linewidth]{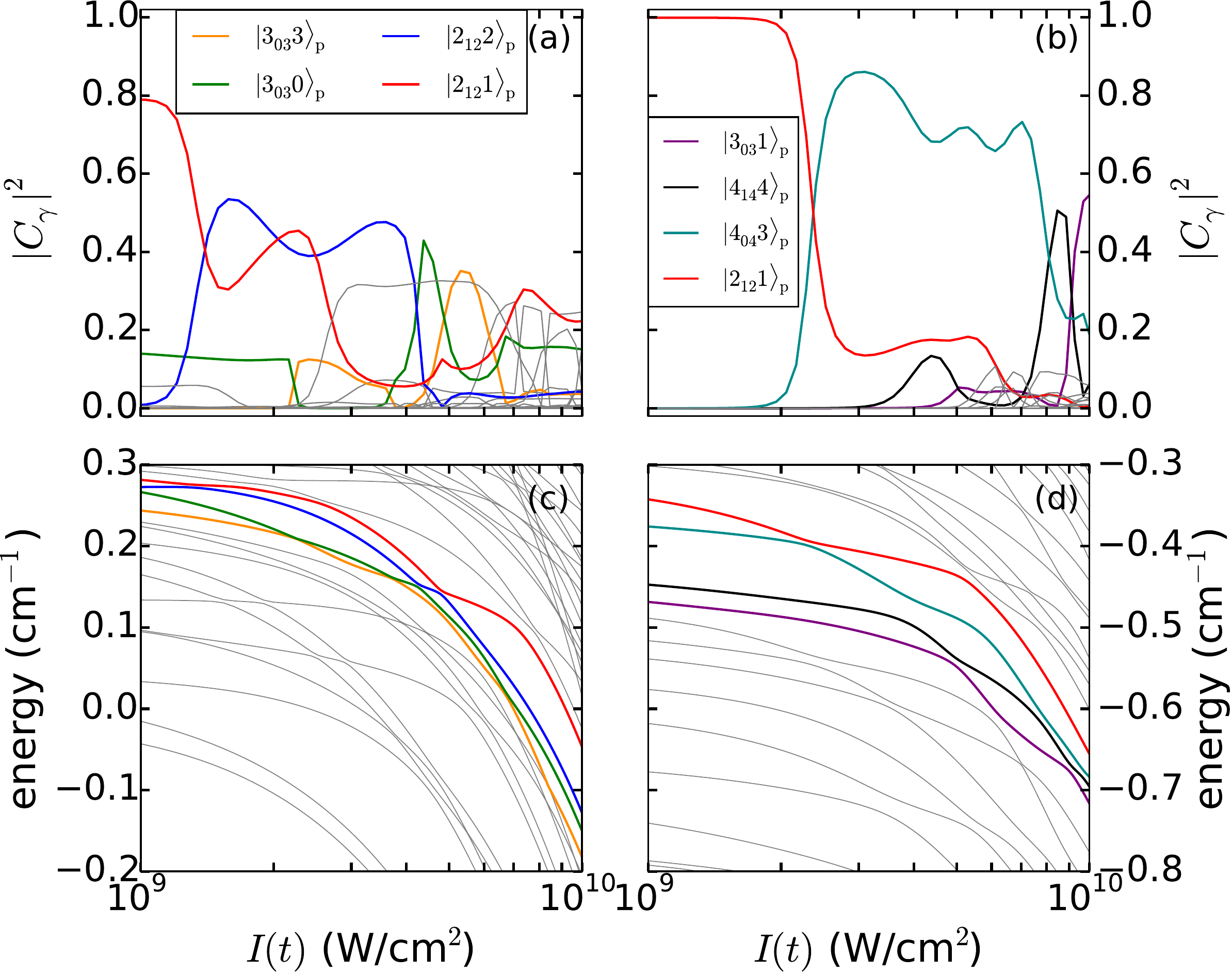}
   \caption{The squares of the projection of the time-dependent wave function onto the
      adiabatic-pendular-state basis of the state $|2_{1 2}1 \rangle_\text{t}$, for a linearly
      polarized ac field through the field-strength regime
      $1.0\times10^9$~\wcmcm~$\leq I(t) \leq $~$1.0\times10^{10}$~\wcmcm and dc field strengths of
      (a) $\Estat=571$~\vcm and (b) $\Estat=20$~\kvcm. Energy level structure for the same
      intensities and dc field strengths (c) $\Estat=571$~\vcm and (d) $\Estat=20$~\kvcm. The angle
      between the ac and dc fields is $\beta=\degree{57.1}$.}
   \label{fig:dynamics_lin}
\end{figure}

We start by analyzing the dynamics for a linearly polarized laser field. Even at low intensities,
the dynamics is highly complicated. This is illustrated in \autoref{fig:dynamics_lin}~(a) and (b),
which show the squares of the projection of the time-dependent wave function onto the adiabatic
basis formed by the eigenstates of the instantaneous Hamiltonian~\eqref{eq:hamiltonian}. The
dynamics is shown for a linearly polarized laser field through the intensity regime
$10^9~\wcmcm\leq{}I(t)\leq10^{10}$~\wcmcm and both dc field strengths, $\Estat=571$~\vcm and
$\Estat=20$~\kvcm. For this example we chose $\beta=\degree{57.1}$, which corresponds to the angle
between EDM and MPA.
\begin{figure*}
   \includegraphics[width=\textwidth]{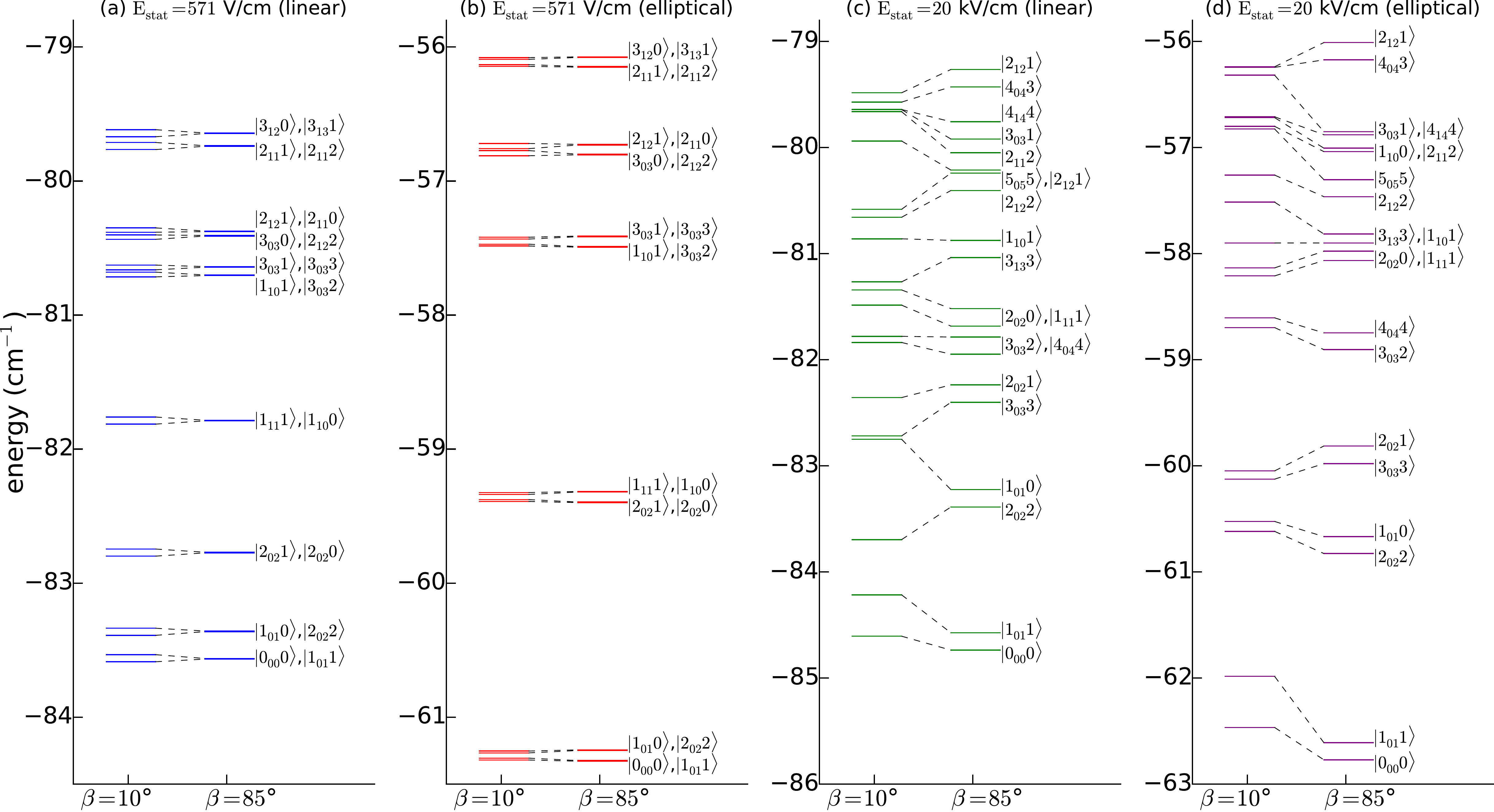}
   \caption{The energies of the 20 lowest lying adiabatic pendular states with even parity under
      reflection on the plane containing the ac and dc fields, at the peak intensity for (a)
      linearly polarized ac field and $\Estat=571$~\vcm, (b) elliptically polarized ac field and
      $\Estat=571$~\vcm, (c) linearly polarized ac field and $\Estat=20$~\kvcm, (d) elliptically
      polarized ac field and $\Estat=20$~\kvcm.}
   \label{fig:energies_peak}
\end{figure*}

For a weak dc field of 571~\vcm, the population of the adiabatic pendular state
$|2_{1 2}1 \rangle_\text{p}$ is already significantly reduced at $1.0\times10^9$~\wcmcm compared to
the laser field-free value of 1.0, see left side of~\autoref{fig:dynamics_lin}~(a). This is due to
population transfer at the splitting of the $|2_{12}M\rangle_\text{p}$ manifold with $M=1,2$
occurring at even lower intensities~\cite{Omiste:PRA94:063408}. In contrast, for $\Estat=20$~\kvcm,
the energy gap between the states within this manifold is so large that their coupling is
significantly reduced, preventing population transfer within the manifold as the ac field is turned
on.

In this low intensity regime, the large number of avoided crossings,
see~\autoref{fig:dynamics_lin}~(c) and (d), is the main source of the non-adiabatic behavior and
leads to many adiabatic pendular states being involved in the dynamics. At each avoided crossing,
the energy spacing among the involved pendular states strongly depends on the dc field strength and
the angle between the ac and dc fields. Thus, the adiabatic pendular states contributing to the
time-dependent wave function significantly vary for the dc field strengths considered here,
\cf~\autoref{fig:dynamics_lin}~(a) and (b). By changing the angle $\beta$ or any other field
parameter, the contributions of the adiabatic pendular states to the time-dependent wave function
also vary strongly (not shown here). We emphasize that avoided crossings play an important role
during the whole rotational dynamics of excited states and are one of the main sources of
non-adiabatic effects. The dynamics through an avoided crossing depends, additionally, on the
temporal profile of the laser. Achieving completely adiabatic dynamics at arbitrary avoided
crossings, \ie, no population transfer between the involved pendular states, would be very
challenging since it requires an extremely slowly increasing intensity due to the extremely narrow
spacing between the involved states; \ie, to achieve adiabatic dynamics corresponding to the energy
and field-strength range presented in~\autoref{fig:dynamics_lin} and~\autoref{fig:dynamics_ell}
microsecond-long laser pulses or a continuous-wave control laser~\cite{Deppe:OE23:28491} would be
necessary.

At stronger laser fields, quasi-degenerate pendular doublets are formed providing an additional
source of non-adiabatic effects. During the doublet formation with increasing $I(t)$, the energy
splitting and the directional properties of the two pendular states change in a way that depends
strongly on the external field parameters, in particular the dc field strength and the angle
$\beta$. Thus, the influence of the doublet formation on the rotational dynamics and the mixed-field
orientation at the peak intensity varies significantly if the field configuration is modified.

\begin{figure}
   \includegraphics[width=\linewidth]{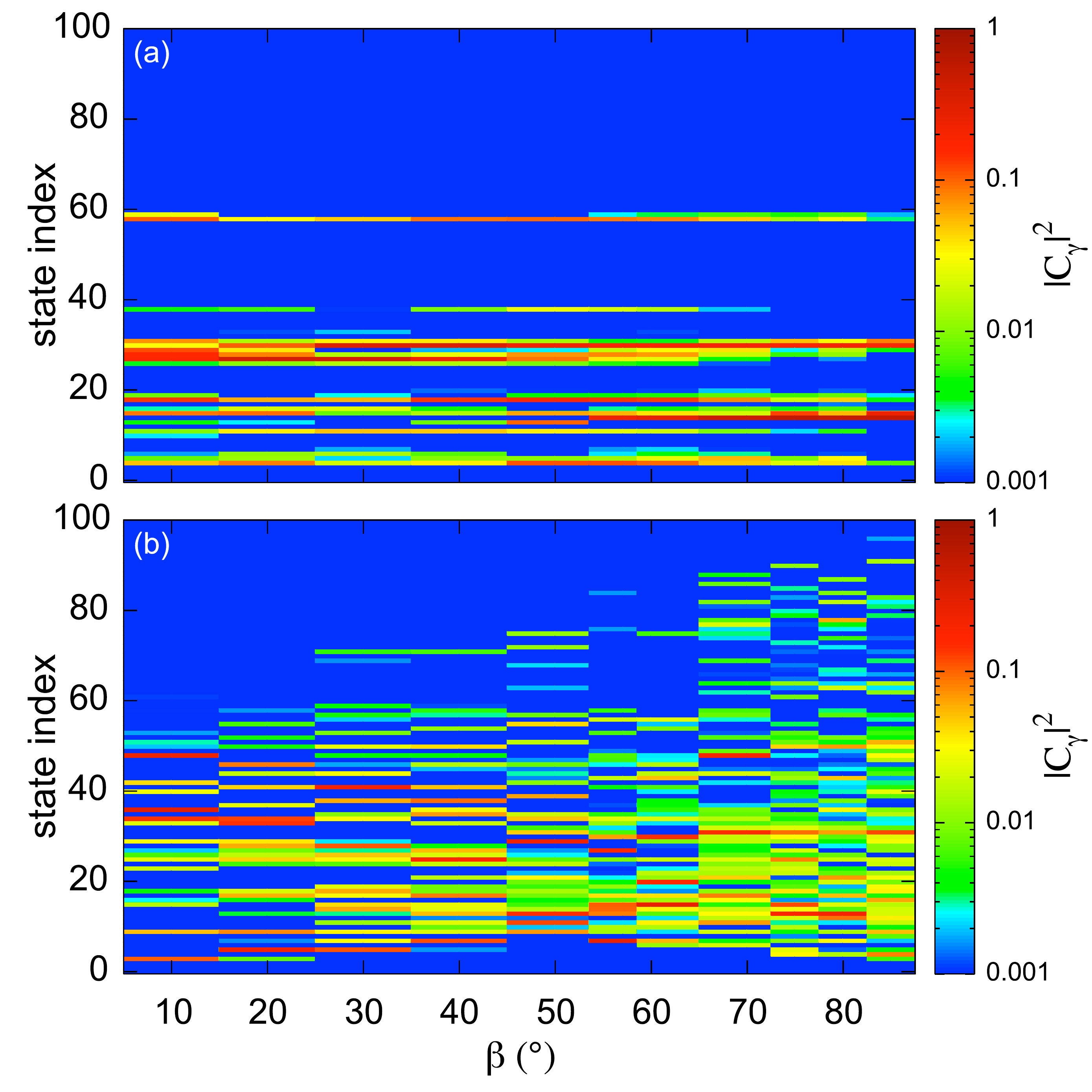}
   \caption{For the state $|2_{1 2}1 \rangle_\text{t}$, the squares of the projection of the wave
      function at the peak of a linearly polarized laser pulse onto the adiabatic pendular states as
      a function of $\beta$ and the index of the pendular states according to energetic order. The
      dc field strength is (a) $\Estat=571$~\vcm and (b) $\Estat=20$~\kvcm. The index of the
      adiabatic pendular state $|2_{1 2}1 \rangle_\text{p}$ is 14 and 19 for the weak and strong dc
      field, respectively.}
   \label{fig:colormap_lin}
\end{figure}

For $\Estat=571$~\vcm, the pendular doublets can be clearly observed for all values of $\beta$,
see~\autoref{fig:energies_peak}~(a), which shows the energies of the 20 lowest lying adiabatic
pendular states at the peak of the laser pulse for $\beta=\degree{10}$ and $\beta=\degree{85}$. The
energy separation within these pendular doublets, $\Delta{E}\sim2\Estat\mu_z\cos\beta$, decreases as
the dc field is rotated towards the perpendicular field configuration. For $\beta=\degree{85}$, the
small energy splitting and rapid formation of the pendular doublets with increasing laser intensity
leads to a redistribution of the population within the quasi-degenerate doublets. This
redistribution of the population for $\beta=\degree{85}$ can be observed in the projection of the
wave function onto the adiabatic basis at the peak intensity, shown
in~\autoref{fig:colormap_lin}~(a), where the two adiabatic pendular states in several doublets have
similar weights. This can also be observed in the field-dressed-dynamics movies in the supplementary
material (see~\autoref{subsection_SI}). Since the quasi-degenerate doublets consist of strongly aligned pendular
states that are oriented in opposite directions along the LFF $Z$-axis, the overall orientation of
the state $|2_{1 2}1 \rangle_\text{t}$ at the peak intensity decreases as $\beta$ approaches
\degree{90}, see~\autoref{fig:orientation_2121}~(a). A similar population redistribution within the
pendular doublets also occurs for other excited states, and, as a result, the orientation of the
thermal ensemble in~\autoref{fig:orientation_thermal}~(a) decreases as $\beta$ approaches
\degree{90}.

As $\beta$ decreases, the increasing $Z$-component of the dc field leads to larger energy splitting
between the two adiabatic pendular states in the doublets. As a result, for small $\beta$ we find
less population transfer among the involved adiabatic pendular states than for $\beta$ close to
\degree{90}. The weights of two adiabatic pendular states forming a quasi-degenerate pair can differ
strongly at the peak intensity for $\beta\lesssim\degree{70}$, \cf~\autoref{fig:colormap_lin}~(a).
In the final population distribution in~\autoref{fig:colormap_lin}~(a), we encounter gaps of
unpopulated pendular states that can be explained by population transfer at avoided crossings
occurring at high intensities. When the pendular doublets are already formed, we encounter avoided
crossings involving four adiabatic pendular states. In these highly non-adiabatic avoided crossings,
the population of the oriented (antioriented) state passes to the oriented (antioriented) state of
the adjacent doublet (see~\autoref{subsection_SI}). Due to this mainly diabatic rotational dynamics at high
intensities only a few adiabatic pendular states, distributed over a wide range of energies,
significantly contribute to the final wave function. These pendular states mainly determine the
orientation at different $\beta$. Depending on the angle between the ac and dc fields, the
contributions of either oriented or antioriented adiabatic states are dominant in the final wave
function. This leads to the oscillating behavior of $\langle\cos\theta_{Zz}\rangle$ versus $\beta$
in~\autoref{fig:orientation_2121}~(a).

For a strong dc field of $\Estat=20$~\kvcm, the arrangement in pendular doublets can still be
observed for $\beta$ close to \degree{90}, where the energy separation within the doublets is much
larger than for a weak dc field, see~\autoref{fig:energies_peak}~(c). For small angles, the lower
lying adiabatic states do not form pendular doublets having opposite orientation, but there are
nearly degenerate adiabatic states oriented in the same direction. For $\beta=\degree{85}$, due to
the large energy separation, the pendular doublet formation does not lead to a significant
population transfer between the two pendular states as for a weak dc field and the same angle. Thus,
the field-dressed dynamics for $\Estat=20$~\kvcm is dominated by avoided crossings between
neighboring levels. The population transfer occurring at these avoided crossings gives rise to a
broad and homogeneous distribution of populated pendular states at the peak intensity,
see~\autoref{fig:colormap_lin}~(b). As a result, we find similar contributions of oriented and
antioriented adiabatic pendular states for $\beta=\degree{85}$, giving rise to a small orientation
along the $Z$-axis of the state $|2_{12}1\rangle_\text{t}$. We find other time-dependent excited
states that are strongly oriented or antioriented at the peak intensity, depending on the weights of
the adiabatic states in their final wave functions. The opposite orientations of these
time-dependent states cancel in the thermal ensemble, which shows a weak orientation for
$\beta=\degree{85}$.

If $\beta$ is small, the initial brute-force orientation of the EDM, induced by the strong dc field,
implies a significant orientation of the MFF $z$-axis along the LFF $Z$-axis before the laser pulse
is turned on. For laser intensities above $1.0\times10^{10}$~\wcmcm, the interaction with the ac
field becomes dominant over the interaction with the dc field and the MPA becomes aligned along the
$Z$-axis. Due to this strong confinement of the MPA, the orientation of many adiabatic pendular
states along the $Z$-axis increases while some pendular states gradually become strongly
antioriented. The change in the $\langle\cos\theta_{Zz}\rangle$ versus the laser intensity can be
observed in the movies provided in the supplementary material (see~\autoref{subsection_SI}). The pendular states
initially becoming antioriented in this intermediate intensity regime are mostly highly excited
states that are not populated so that mainly oriented adiabatic states continue to contribute to the
rotational dynamics in this regime. As the laser intensity increases further, we encounter numerous
avoided crossings between these strongly oriented and strongly antioriented pendular states. At such
an avoided crossing, the involved adiabatic pendular states interchange their directional
properties, \ie, the previously oriented (antioriented) state becomes antioriented (oriented). Since
these avoided crossings are traversed diabatically, the time-dependent wave function does not
acquire population on the antioriented adiabatic pendular states. As a result, the distribution
in~\autoref{fig:colormap_lin}~(b) shows gaps of unpopulated adiabatic states that are mainly
antioriented at the peak intensity. Thus, at the peak intensity, the state
$|2_{1 2}1 \rangle_\text{t}$ is strongly oriented along the LFF $Z$-axis for small values of
$\beta$, see~\autoref{fig:orientation_2121}~(a). The dynamics of other excited states shows a
similar behavior, thus enhancing the orientation of the thermal ensemble as seen
in~\autoref{fig:orientation_thermal}~(a).

For $\beta$ close to \degree{90}, \ie, small angles between the dc field and the LFF $Y$-axis, the
initial brute-force orientation of the EDM implies an orientation of the MFF $y$-axis along the LFF
$Y$-axis before the laser pulse is turned on. This orientation is maintained for several adiabatic
pendular states even at the peak intensity but we also find other pendular states being weakly
antioriented or not oriented (see~\autoref{subsection_SI}). This leads to an overall moderate orientation of the MFF
$y$-axis along the LFF $Y$-axis for the state $|2_{1 2}1 \rangle_\text{t}$ as well as the thermal
average, for $\beta$ close to \degree{90}, see~\autoref{fig:orientation_2121}~(b)
and~\autoref{fig:orientation_thermal}~(b). If $\beta$ is small, the dc electric field does not
induce an initial orientation of the MFF $y$-axis along the LFF $Y$-axis. In the presence of the
laser field, different time-dependent excited states may become moderately oriented in opposite
directions or show no orientation along the LFF $Y$-axis at the peak intensity. Thus, the thermal
ensemble shows a very small orientation $\langle\cos\theta_{Yy}\rangle_T$ for small $\beta$.

\subsection{Elliptically polarized laser}
We now consider the rotational dynamics in elliptically polarized laser fields. In the absence of
the dc field, the elliptically polarized ac field induces 3D alignment of the CPC molecules, with
the $z$ and $y$ molecular axes aligned along the LFF $Z$ and $Y$-axes, respectively. Compared to the
linearly polarized cases analyzed in the previous section, and since the total intensities for both
polarizations are equal, the effective intensities of the elliptically polarized laser field along
these two axes are weaker. This implies a smaller alignment along the $Y$-axis, as is illustrated
in~\autoref{fig:alignment}. In this section, we analyze how this change in the ac field polarization
affects the rotational dynamics and mixed-field orientation.

\begin{figure}
   \includegraphics[width=\linewidth]{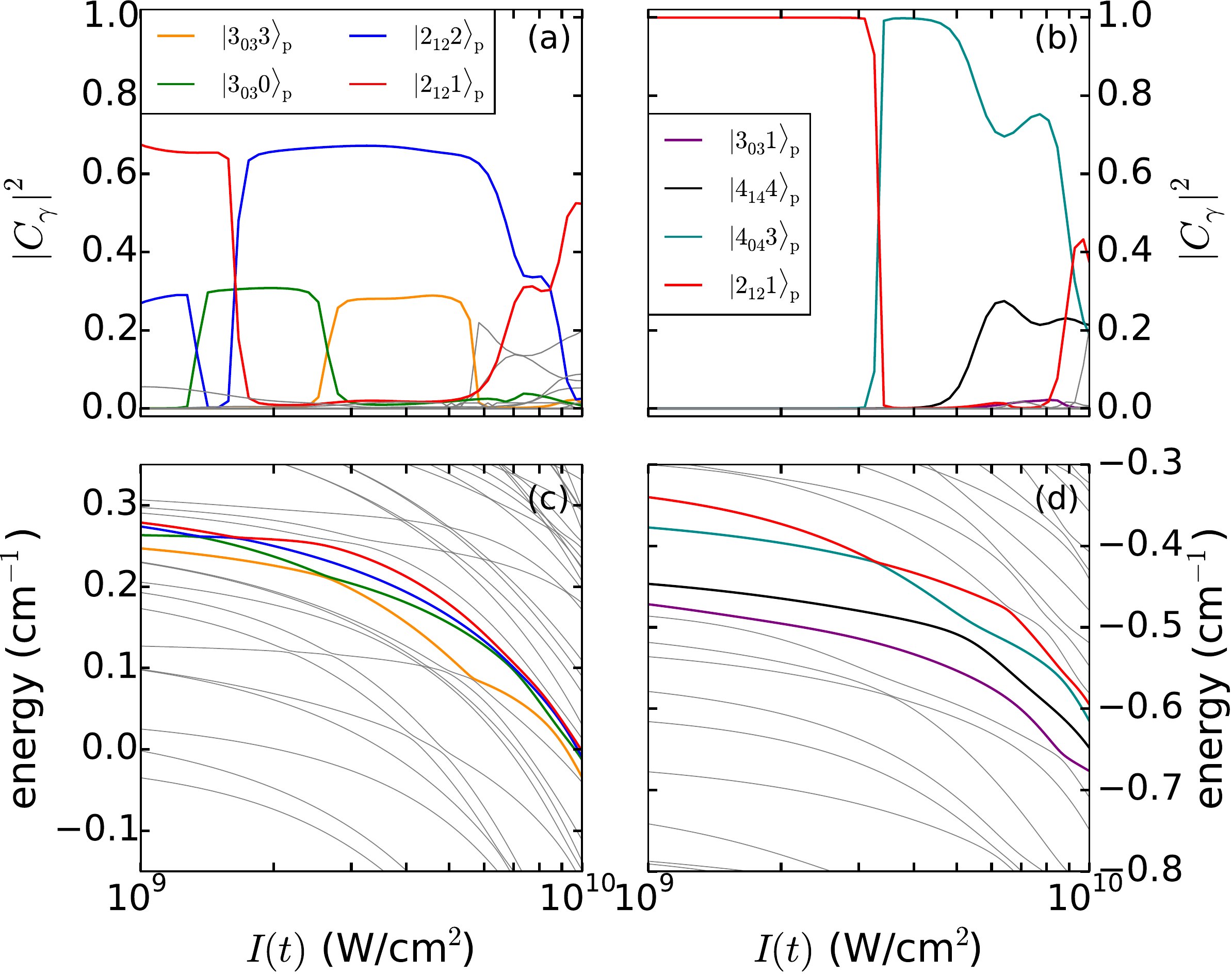}
   \caption{The squares of the projection of the time-dependent wave function onto the
      adiabatic-pendular-state basis of the state $|2_{1 2}1 \rangle_\text{t}$, for an elliptically
      polarized ac field through the field-strength regime
      $1.0\times10^{9}$~\wcmcm~$\leq I(t) \leq $~$1.0\times10^{10}$~\wcmcm and dc field strengths of
      (a) $\Estat=571$~\vcm and (b) $\Estat=20$~\kvcm. Energy level structure for the same
      intensities and dc field strengths (c) $\Estat=571$~\vcm and (d) $\Estat=20$~\kvcm. The angle
      between the ac and dc fields is $\beta=\degree{57.1}$.}
   \label{fig:dynamics_ell}
\end{figure}

For low intensities, $10^9~\wcmcm\leq{}I(t)\leq~10^{10}$~\wcmcm, the squares of the projection of
the time-dependent wave function for the state $|2_{1 2}1 \rangle_\text{t}$ are shown
in~\autoref{fig:dynamics_ell}~(a) and (b) for $\beta=\degree{57.1}$ and $\Estat=571$~\vcm and
$\Estat=20$~\kvcm, respectively. In this intensity regime, we find a similarly complex dynamics as
for a linearly polarized laser field presented in~\autoref{fig:dynamics_lin}. For
\mbox{$\Estat=571$~\vcm} the splitting of the $|2_{12}M\rangle_\text{p}$ manifold at weaker
intensities leads to population transfer among the two pendular states within this manifold. For a
strong dc field with $\Estat=20$~\kvcm, this effect is not significant and the population of the
state $|2_{1 2}1 \rangle_\text{p}$, is still approximately 1.0 at $1.0\times10^{9}$~\wcmcm. The
large number of avoided crossings in~\autoref{fig:dynamics_ell}~(c) and (d) dominate the rotational
dynamics, which involves several adiabatic pendular states even for low intensities. Differences
between the dynamics for linearly and elliptically polarized ac fields gain importance at higher
intensities where the elliptically polarized ac field significantly influences the energy level
structure and the directional features of the adiabatic pendular states.
\begin{figure}
   \includegraphics[width=\linewidth]{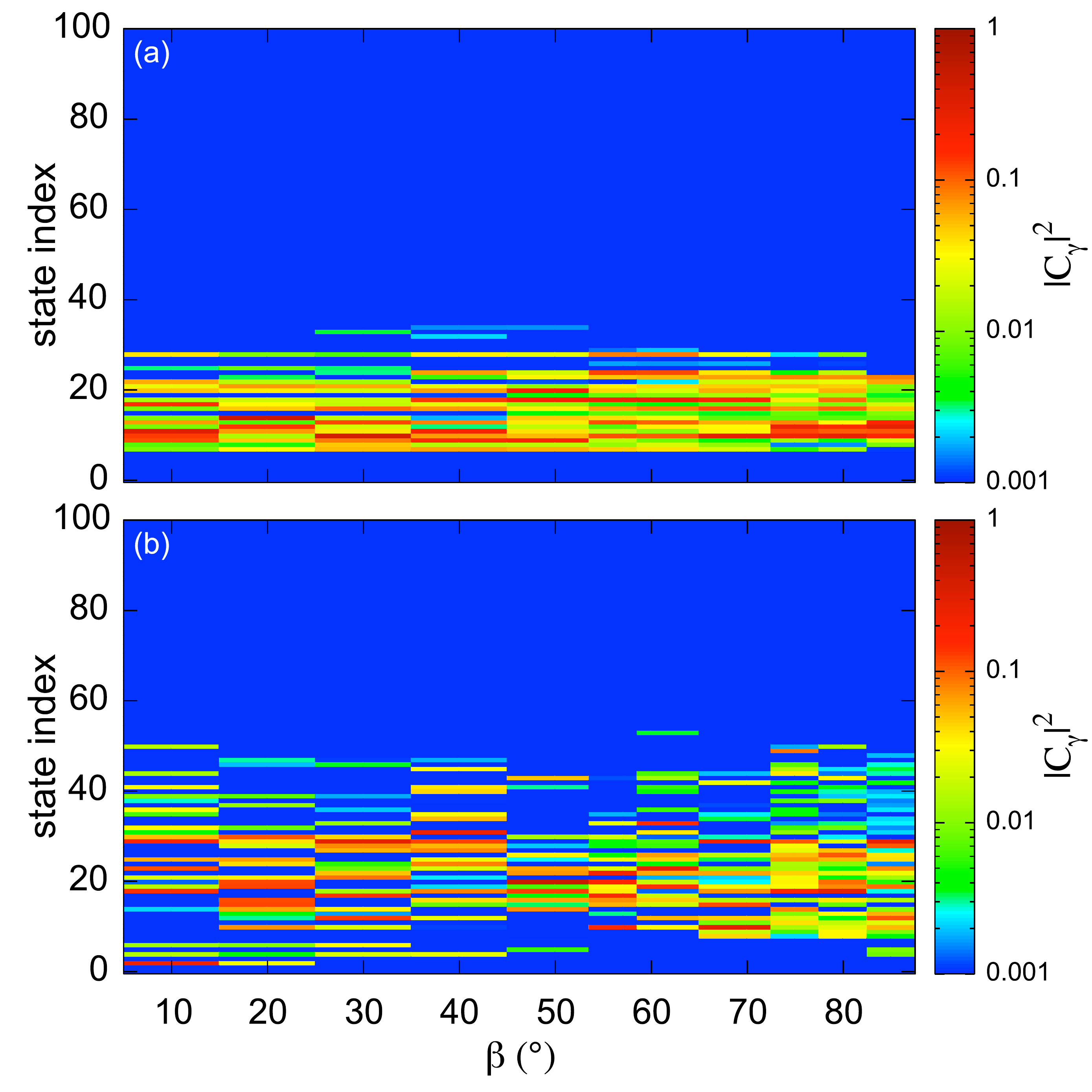}
   \caption{For the state $|2_{1 2}1 \rangle_\text{t}$, the squares of the projection of the wave
      function at the peak of a elliptically polarized laser pulse onto the adiabatic pendular
      states as a function of $\beta$ and the index of the pendular states according to energetic
      order. The dc field strength is (a) $\Estat=571$~\vcm and (b) $\Estat=20$~\kvcm. The index of
      the adiabatic pendular state $|2_{1 2}1 \rangle_\text{p}$ is 14 and 19 for the weak and strong
      dc field, respectively.}
   \label{fig:colormap_ell}
\end{figure}

Analogously to the linearly polarized ac field, near-degenerate pendular doublets are formed as the
laser intensity increases. For $\Estat=571$~\vcm and \mbox{$I(t)\gtrsim10^{11}$~\wcmcm}, we
additionally encounter the formation of near-degenerate quadruplets, after the doublets are already
formed. This arrangement in doublets and quadruplets can be observed in the energies of the
adiabatic pendular states at the peak intensity, shown in~\autoref{fig:energies_peak}~(b). The
energy separations between consecutive quadruplets are larger than between consecutive doublets in
the case of a linearly polarized ac field. Due to the quadruplet formation for the elliptically
polarized laser field, we find a smaller number of avoided crossings between near-degenerate groups
of pendular states than for linear polarization. These features of the energy level structure limit
the population transfer to highly excited states and the population at the peak intensity is
confined to a small region of relatively low-lying adiabatic pendular states for all values of
$\beta$, \cf~\autoref{fig:colormap_ell}~(a) and~\autoref{subsection_SI}.

For the weak dc field and $\beta$ close to \degree{90}, a redistribution of the population occurs
during the formation of the near-degenerate doublets similar to the case of a linearly polarized ac
field (see~\autoref{subsection_SI}). This is reflected in the weights of the adiabatic pendular states contributing to the
time-dependent wave function at the peak intensity, \cf~\autoref{fig:colormap_ell}~(a), where there
are several neighboring adiabatic levels with similar contributions. This redistribution leads to
the decrease of $\langle\cos\theta_{Zz}\rangle$ for $\beta\gtrsim\degree{75}$ for the state
$|2_{1 2}1 \rangle_\text{t}$ in~\autoref{fig:orientation_2121}~(a) and the thermal ensemble
in~\autoref{fig:orientation_thermal}~(a), similar to the case of a linearly polarized laser pulse.
For $\beta=\degree{85}$, the states forming each of the two doublets in a near-degenerate quadruplet
are oriented in opposite directions along the LFF $Z$-axis, but in the same direction along the
$Y$-axis (see~\autoref{subsection_SI}). Thus, the state $|2_{1 2}1 \rangle_\text{t}$ shows a moderate orientation along the
$Y$-axis at this angle see~\autoref{fig:orientation_2121}~(b). We find other time-dependent excited
states that are antioriented at the peak intensity due to different adiabatic pendular states
contributing to their rotational dynamics. In the thermal ensemble, the contributions of states,
which are oriented along the $Y$-axis, become more dominant as $\beta$ approaches \degree{90},
giving rise to the increase of $\langle\cos\theta_{Yy}\rangle$ for increasing $\beta$
in~\autoref{fig:orientation_thermal}~(b).

For a strong dc electric field, pendular doublets can be observed for $\beta$ close to \degree{90}
and some low-lying pendular states for small $\beta$. Near-degenerate quadruplets are not formed,
see~\autoref{fig:energies_peak}~(d). Thus, avoided crossings are the main source of non-adiabatic
phenomena in the rotational dynamics at higher intensities.

For $\beta=\degree{10}$, a qualitatively similar rotational dynamics can be observed for both,
linear and elliptical, polarizations of the ac field, but with different adiabatic states
contributing to the time-dependent wave functions (see~\autoref{subsection_SI}). The distributions of the population of adiabatic
pendular states in the wave function at the peak intensity show the same features, \ie, the
approximate number and range of contributing pendular states, in both cases,
see~\autoref{fig:colormap_ell}~(b) and~\autoref{fig:colormap_lin}~(b). The state
$|2_{1 2}1 \rangle_\text{t}$ also shows a significant orientation of the MFF $z$-axis along the LFF
$Z$-axis at the peak intensity of the elliptically polarized ac field,
see~\autoref{fig:orientation_2121}~(a). However, since the weights of the adiabatic pendular states
vary when the ac field polarization is changed, $\langle\cos\theta_{Zz}\rangle$ at
$\beta=\degree{10}$ differs for the linearly and elliptically polarized laser pulses.

Regarding the orientation along the LFF $Y$-axis for the strong dc field and the elliptically
polarized ac field, the state $|2_{12}1\rangle_\text{t}$ shows an analogous behavior for
$\beta=\degree{85}$ as for the orientation along the LFF $Z$-axis for $\beta=\degree{10}$. Once the
ac field component along the LFF $Y$-axis is sufficiently strong to align the MFF $y$-axis along the
LFF $Y$-axis, the initially brute-force-oriented adiabatic pendular states become either strongly
oriented or strongly antioriented along this axis (see~\autoref{subsection_SI}). As this occurs, several pendular states becoming
antioriented along the $Y$-axis have significant weights in the time-dependent wave function. As a
result, the state $|2_{12}1\rangle_\text{t}$ only shows a moderate orientation of the MFF $y$-axis
along the LFF $Y$-axis at the peak intensity, \cf~\autoref{fig:orientation_2121}~(b). Other excited
states may even be weakly antioriented. Thus, for the thermal ensemble,
$\langle\cos\theta_{Yy}\rangle_T$ in~\autoref{fig:orientation_thermal}~(b) is only moderately
enhanced by the strong dc field.

To summarize, even if the field-dressed dynamics is highly non-adiabatic, the directional features
of individual excited rotational states and, consequently, of the ensemble average, can be
controlled to some extend. A significant orientation of the MFF $z$-axis along the LFF $Z$-axis can
be achieved by applying a strong dc field for small values of $\beta$ and both a linearly and
elliptically polarized laser field. An analogous enhancement of the orientation of the MFF $y$-axis
along the LFF $Y$-axis is obtained for an elliptically polarized ac field and $\beta$ close to
\degree{90}.

\section{Conclusion}
We present a time-dependent analysis of the rotational dynamics of a non-symmetric molecule in
combined ac and dc electric fields. We investigate the influence of the dc field strength and the
angle between the ac and dc fields during the turn-on of linearly and elliptically polarized laser
pulses. Our theoretical study is focused on the prototypical CPC molecule, which has a permanent
dipole moment with a direction that is neither parallel to any principle axis of inertia nor to any
principle axis of polarizability.

Our computational results for a thermal ensemble at 200~mK agree quantitatively with
mixed-field-orientation experiments of state-selected CPC molecules~\cite{Hansen:JCP139:234313},
including the 1D orientation along the $Z$-axis, and the smooth $\beta$-dependence of the
orientation that could not be explained by the previous adiabatic description. Our time-dependent
description also shows that in the experiment 3D orientation was achieved for the employed
combination of the elliptically polarized laser pulse and the weak dc field.

The analysis of the time-dependent dynamics of individual states, which we described for the
prototypical state $|2_{12}1\rangle_\text{t}$, shows highly non-adiabatic dynamics. These
non-adiabaticities arise due to avoided crossings, $M$-manifold splitting, and the formation of
pendular doublets and quadruplets. The relevance of these phenomena and the exact dynamics are
highly dependent on the field parameters, \eg, the laser pulse, the dc field strength, and the angle
between the two fields. Generally, the distribution of population over many adiabatic pendular
states leads to a weak net orientation, because \emph{a priori} as many oriented as antioriented
pendular states are involved in the field-dressed dynamics. However, for strong dc electric fields,
preferred contributions of oriented pendular states to the wave function are obtained and strong
degrees of orientation are achieved even for rotationally excited states.

For a ground-state ensemble, which for complex molecules can be produced by state-specific
guiding~\cite{Putzke:PCCP13:18962}, adiabatic 3D orientation can be achieved by combining a long
laser pulse with a moderately strong dc electric field. For excited states, although the dynamics is
non-adiabatic, their directional features can be controlled to some degree by applying strong dc
fields. Furthermore, we show that 3D orientation of a thermal ensemble can be achieved in two ways:
Firstly, by combining a linearly polarized laser field with a strong dc electric field, effectively
locking the MPA to the laser polarization axis and the EDM to the dc electric field direction.
Secondly, by combining an elliptically polarized laser field with a strong dc electric field to
induce mixed-field orientation of the 3D aligned sample.

While the current work is focused on the prototypical complex molecule CPC, similar rotational
dynamics are expected for other non-symmetric molecules. Even if the principle axes of inertia and
polarizability are non-parallel, mixed-field orientation is expected. Since the non-adiabatic
effects strongly depend on the molecular properties, accurate predictions of the results of
mixed-field experiments requires an accurate description, e.g., a TDSE analysis of the rotational
dynamics. The present work paves the way to an accurate description of the mixed-field orientation
of large molecules in general, which is highly relevant for molecular-frame imaging experiments of
complex molecules, where experimental determinations of the degree of alignment and orientation
become very difficult and corresponding simulations of the angular
control are necessary to analyze the experimental data~\cite{Filsinger:PCCP13:2076,
   Barty:ARPC64:415}.

\section{Supplementary Material}
\label{subsection_SI}
See supplementary material for movies of the field-dressed rotational dynamics of
the state $|2_{12}1\protect\rangle_\text{t}$ and the orientation cosines of the adiabatic
pendular states.

\section{Acknowledgements}
We gratefully acknowledge Juan J.\ Omiste for developing the computer code to solve the
time-dependent Schrödinger equation.

In addition to DESY, this work has been supported by the Helmholtz Association ``Initiative and
Networking Fund'', by the \emph{Deutsche Forschungsgemeinschaft} (DFG) through the excellence
cluster ``The Hamburg Center for Ultrafast Imaging -- Structure, Dynamics and Control of Matter at
the Atomic Scale'' (CUI, EXC1074), and by the European Research Council through the Consolidator
Grant COMOTION (ERC-Küpper-614507). R.G.F.\ gratefully acknowledges financial support by the Spanish
project FIS2014-54497-P (MINECO) and by the Andalusian research group FQM-207 and the grant
P11-FQM-7276.

\bibliography{string,cmi}
\end{document}